\documentclass[iop,apj]{emulateapj}

\renewcommand{\figurename}{Figure}

\newcommand{\del}[2]{\frac{\partial #1}{\partial #2}}
\newcommand{\deld}[2]{\frac{\partial^2 #1}{\partial #2^2}}

\newcommand{\unit}[1]{\,\mathrm{#1}}

\newcommand{\figref}[1]{\figurename ~\ref{fig:#1}}

\renewcommand{\bibname}{Reference}
\makeatletter
\def\ddigit#1{\setbox0=\hbox{9}\setbox1=\hbox{#1}%
  \ifdim \wd0<\wd1 #1\else 0#1\fi}
\newcounter{hour}
\newcounter{HOUR}
\newcounter{minuite}
 \divide \c@hour by 60
 \multiply \c@HOUR by 60
 \advance \c@minuite by -\c@HOUR
\def\digitalhour{\ddigit{\thehour}}
\def\digitalminuite{\ddigit{\theminuite}}

\newcommand{\printtime}{\the\year /\two@digits\month /\two@digits\day
~~\dayofweek~~\digitalhour:\digitalminuite}
\makeatother
\usepackage{natbib, aas_macros}
\usepackage{amsmath}
\usepackage{subfigure}
\usepackage{graphicx}
\usepackage[right,pagewise,mathlines]{lineno}
\usepackage[usenames]{color}
\graphicspath{{./images/}}

\shorttitle{STABILITY OF COSMIC RAY MODIFIED SHOCKS}
\shortauthors{SAITO, HOSHINO \& AMANO}

\begin{document}

\title{Stability of Cosmic Ray Modified Shocks: Two-Fluid Approach}

\author{TATSUHIKO SAITO\altaffilmark{1}, MASAHIRO HOSHINO, and TAKANOBU AMANO}
\affil{Department of Earth and Planetary Science, The University of
Tokyo, Tokyo 113-0033, Japan; saito-t@eps.s.u-tokyo.ac.jp}

\begin{abstract}
%#!platex main.tex

%%%%%%%%%%%%%%%%%%%%%%%%%%%%%%%%%%%%%%%%%%%%%%%%%%%%%%%%%%%%%%%%%%%%%%
The stability of 
the cosmic ray modified shock (CRMS) is studied by means
of numerical simulations. Owing to the nonlinear feedback of cosmic-ray
(CR) acceleration, a downstream state of the modified shock can no 
longer be uniquely
determined for given upstream parameters. It is known that up to three
distinct solutions exist, which are characterized by the CR production
efficiency as ``efficient'', ``intermediate'' and ``inefficient''
branches. The stability of these solutions is investigated by performing
direct time-dependent simulations of a two-fluid model. It is found that both
the efficient and inefficient branches are stable even against a
large-amplitude 
perturbation, while the intermediate one is always unstable and evolves into
the inefficient state as a result of nonlinear time development. This bistable
feature is robust in a wide range of parameters and does not depend on the
injection model. Fully nonlinear time evolution of a
hydrodynamic shock with injection  results
in the least efficient state in terms of the CR production, consistent with
the bistable feature. This suggests that the CR production efficiency at
supernova remnant shocks may be lower than previously discussed in the
framework of the 
nonlinear shock acceleration theory considering the
efficient solution of the CRMS.

%%% Local Variables:
%%% mode: yatex
%%% TeX-master: "main.tex"
%%% End:
\end{abstract}


%#!platex main.tex

\section{INTRODUCTION}\label{sec:intro}
%%%%%%%%%%%%%%%%%%%%%%%%%%%%%%%%%%%%%%%%%%%%%%%%%%
From many ground-based and satellite observations, it is now widely believed
that cosmic rays (CRs) with energies at least up to the knee $(\sim 10^{15.5}
\unit{eV})$, are accelerated by shock waves of supernova remnants (SNRs).
Diffusive Shock Acceleration (DSA), or the first order Fermi acceleration,
which has been the standard theory since the late 1970s
\citep{1978MNRAS.182..147B,1978ApJ...221L..29B,1977ICRC...11..132A}, predicts
a power-law type spectrum of accelerated particles with its index solely
determined by the shock compression ratio. The index approaches  the
universal value of 2, which is more or less consistent with observations of
the local cosmic-ray spectrum $\sim 2.7$ if the propagation effect is taken
into account. It has also been supported by more detailed observations of
shocks in various environments. Since the accelerated particles diffuse ahead
of the shock, one would expect the formation of a precursor region in which
energetic particle intensity gradually increases toward the shock. X-ray
observations of SNR shocks as well as in-situ observations of shocks in the
heliosphere are consistent with this picture.
\citep{2003ApJ...589..827B,2005ApJ...621..793B,1999Ap&SS.264..481S,2005JGRA..11009S12T,2006AdSpR..37.1408T}.

Although the DSA theory was initially constructed in the test-particle limit,
it turns out to be a very efficient acceleration mechanism. Therefore, a lot
of work has been devoted to clarify the role of possible feedback effects
from the accelerated particles.
Namely, once the energy density of accelerated particles becomes comparable to
that of the background plasma, their back-reaction may substantially modify
the shock structure itself, which then affects the spectrum of the accelerated
particles. Such a nonlinear shock modified by the presence of the accelerated
particles is called a cosmic ray modified shock (CRMS)
\citep{1981ApJ...248..344D,1983RPPh...46..973D,1982A&A...111..317A}.

%One notable effect caused by the shock modification appears in the spectrum of
%energetic particles. 
%Since the acceleration of CRs effectively increases the
%degrees of freedom of the thermal gas component, the total shock compression
%ratio becomes larger than that of the hydrodynamic shock.  
%In DSA theory in
%the test particle limit, the power-law index 
%$s$ of the energy spectrum is given by $s = 
%(r+2)/(r-1)$, where $r$ is the shock compression ratio. The shock modification
%therefore makes the spectrum even harder, indicating a positive feedback. 
%In reality, the CR pressure gradient exerted to the background plasma in the
%precursor region decelerates the flow and the plasma is somewhat
%compressed. The plasma in the precursor is further compressed across a
%subshock, which is a discontinuous transition connecting the precursor and the
%downstream state. 
%Relatively low-energy CRs diffusing around the subshock
%whose mean free paths are smaller than the typical scale length of the
%precursor see a compression ratio smaller than the total, leading to a softer
%spectrum in this energy range. Consequently, the energy spectrum in the whole
%energy range may become concave. 
It has been known that the energy spectrum of CRs become concave because
of the deceleration of the upstream flow in the precursor region.
This spectral characteristic 
%due to the strong shock modification 
may explain recent X-ray and $\gamma$-ray observations of
young SNRs \citep{2006ApJ...648L..33V,2012A&A...538A..81M}. It is also noted
that the characteristic deceleration of the flow ahead of strong
interplanetary shocks has been reported
\citep{1999Ap&SS.264..481S,2005JGRA..11009S12T}.

Direct observational identification of strongly modified shocks is still,
however, a controversial issue. \cite{2009Sci...325..719H} estimated a
downstream proton temperature of $\sim 2.3 \unit{keV}$ from H$\alpha$ 
observations of RCW 86, which is more than one order of magnitude
smaller than would be expected from the standard Rankine-Hugoniot
relations ($\sim 42 \unit{keV}$). They then concluded that the downstream CR
energy may exceed  half of the total energy. Similar observations suggesting
efficient CR production have also been reported
\citep{2000ApJ...543L..61H,2000ApJ...543L..57D,2010ApJ...719L.140H,2005ApJ...634..376W,Cassam-Chena_2008ApJ_680_1180C}. 
On the
other hand, \cite{2013arXiv1304.1261F} estimated, 
using observations of $\gamma$-rays and interstellar
molecular clouds, that the CR proton energy is only $\sim 0.1 \%$ of the total
kinetic energy in young SNRs RX J1713.7-3946 and RX J0852.0-4622. In the
heliosphere, the energy densities of particles accelerated by shocks driven by
coronal mass ejections (CMEs) are estimated using in-situ observations. The
results indicate that the energetic particles account for at most $10-20 \%$
of the CME kinetic energy, which is not in a strongly modified regime
\citep{2005JGRA..110.9S18M,2006SSRv..124..303M}.

It is well known that one of the most critical issues in the DSA theory is the
maximum attainable energy through this process. \cite{1983A&A...125..249L}
estimated the maximum energy using typical values of SNRs which was found to
be $10^{14} \unit{eV}$ even in the most optimistic scenario; still an order of
magnitude smaller than the knee energy ($\sim 10^{15.5} \unit{eV}$). The
nonlinear back-reaction of CRs may play a key role in resolving the problem. It
becomes apparent from recent high resolution X-ray observations that the
magnetic fields of young SNRs are substantially amplified up to a
$\mathrm{mG}$ level from a typical interstellar value of a few $\mu\mathrm{G}$
\citep{2003ApJ...584..758V,2004A&A...416..595Y,2004A&A...419L..27B,2006AdSpR..37.1902B,2007Natur.449..576U}.
Various mechanisms have been proposed to explain the amplification so far; one
of which indeed attributes this to the back-reaction of accelerated CRs.
\citep[e.g.,][]{1992ApJ...385..193K,2009ApJ...692.1571M,1986MNRAS.223..353D,2012MNRAS.427.2308D}.
It is well known that CRs diffusing ahead of a shock front may drive various
plasma instabilities in the precursor. If the acceleration of CRs is so
efficient that their energy density becomes comparable to the background
plasma, these instabilities may amplify a seed magnetic field in the upstream
more than 100 times. The magnetic field amplification could reduce the mean
free path of CRs thereby increasing the acceleration efficiency. For instance,
according to the estimate by \cite{2004MNRAS.353..550B}, 
the maximum energy goes well beyond
the knee energy. \nocite{2004MNRAS.353..550B}

A peculiar feature of a CRMS is that it possibly has multiple steady-state
solutions, i.e., the downstream state cannot be uniquely determined from given
upstream parameters. This fact was first pointed out by
\cite{1981ApJ...248..344D} by using a two-fluid model in which CRs are
approximated to a massless fluid that interacts with the background plasma
through their pressure. 
\cite{2001ApJ...546..429B} investigated the exact analytical conditions for the
existence of these multiple solutions depending on the Mach
number, the specific heat ratio of the background plasma and CRs in the
two-fluid model.
This model was extended to include the
effect of injection \citep{1993ApJ...406...67Z}, magnetic fields
\citep{1986A&A...160..335W}, 
and to a fully kinetic
treatment in which the diffusion-convection equation 
for CRs and the hydrodynamic 
 equations for the background plasma are coupled with each other
\citep{1997ApJ...485..638M,1997ApJ...491..584M,1996ApJ...473..347M,2001RPPh...64..429M,2005MNRAS.361..907B,2008MNRAS.385.1946A,2009ApJ...694..951R}. Although
the detailed structure of solutions depends on the model, they all possess up
to three distinct solutions in some regions in parameter space, indicating
that this is a generic feature of the nonlinear shock. The three solutions
may be called ``efficient'', ``intermediate'', and ``inefficient'' in terms of
their corresponding CR production efficiencies (see Section \ref{sec:1d_stab}
for details). 
A question naturally arises as to which of these solutions 
indeed exist in nature as the time-asymptotic state 
of a nonlinear particle-accelerating shock.
It is particularly important because the problem
is intimately linked to the maximum energy attainable through the acceleration
process in the efficient branch as well as the CR scenario of magnetic field
amplification. Understanding the stability of these multiple solutions  is thus
crucial for modeling broadband spectra of astrophysical shocks, from which
physical parameters of the acceleration sites can be deduced.

In the original paper of \cite{1981ApJ...248..344D}, they suggested the
possibility of intermediate branch being unstable and these three
branches may have a ``bistable'' feature. 
They conjectured that when the downstream CRs increases (decreases) 
 from the intermediate branch, a self-induced increase
(decrease) may bring the solution toward the efficient (inefficient)
branch.
We note 
that the intermediate branch was previously shown to be
``corrugative'' unstable against perturbations transverse to the shock
\citep{1998A&A...332..385M}.
\cite{1994ApJ...424..263D} conducted time-dependent numerical simulations
adopting the two-fluid model, 
and confirmed that the inefficient and the
efficient branches exist at least as the time-asymptotic states.
Particularly for the efficient branch, it is known that the acoustic
instability occur in the precursor region, and  analytical as well as 
numerical studies on this instability have been given so far 
\citep{1986MNRAS.223..353D,1993ApJ...405..199R,2012MNRAS.427.2308D}.
Nevertheless, to the authors knowledge, a comprehensive investigation of 
the stability of these multiple solutions has not been given 
even within the framework of the two-fluid model.

In the present paper, we study the stability
of the {\it global} CRMS structure 
 in various parameter regimes by means of one-dimensional (1D) direct
time-dependent numerical simulations of the two-fluid equations. It is found
that the solutions, both on the efficient and inefficient branches, are stable,
while those on the intermediate branch 
are always unstable, even in 1D (i.e., even
in the
absence of the corrugative instability). Extensive parameter survey
demonstrates that this basic property does not depend on Mach numbers,
fractions of pre-existing CRs, or the injection model and its efficiency. 
We find that solutions, which are initially on the unstable 
intermediate branch, transit to the inefficient ones 
as a result of self-consistent time
evolution. Moreover, the efficient and inefficient solutions are found to be
stable even against large-amplitude perturbations. 
Because of this, self-consistent
time evolution from a hydrodynamic shock always results 
in the least efficient state for given parameters of the
shock. This implies that 
the CR acceleration efficiency 
by an astrophysical shock may not 
necessarily be high as discussed previously in the
context of nonlinear shock acceleration theory.

This paper is organized as follows. In Section \ref{sec:1d_stab}, we describe
the two-fluid model used in the present study and briefly review the
characteristics of the steady-state solutions of CRMSs, both with and without
injection. In Section \ref{sec:1d_hd_stability}, we present simulation results
clarifying the stability of these solutions. Simulations with and without
injection, and simulations with large-amplitude perturbations, including 
time evolution from a gas dynamic shock with injection, are
presented. In Section \ref{sec:sum}, we summarize the results and discuss 
some implications for CR accelerations in astrophysical shocks.

%%% Local Variables:
%%% mode: yatex
%%% TeX-master: "main.tex"
%%% End:
%#!platex main.tex

\section{MODEL} \label{sec:1d_stab}

\subsection{Two-Fluid Model}\label{sec:basic_eqs}

The CRMS was firstly studied by using a two-fluid model proposed by the
seminal paper of \cite{1981ApJ...248..344D}. In this model, both the
background thermal plasma and CRs are approximated as fluids coupled with each
other. The model was later extended to include the effect of injection by
\cite{1993ApJ...406...67Z} with its efficiency expressed by $\alpha$ as
described in the next subsection. The basic equations in this case for a 1D
parallel shock, which is sufficient to capture much of the essential
physics, 
are thus given as follows,
\begin{align}
& \del{\rho}{t}   +\del{}{x}(\rho u)=0, \label{eq:hd_mass}\\
& \del{}{t}(\rho u)+ \del{}{x}(\rho u^{2}+p_{g}+p_{c})=0, \label{eq:hd_mom}\\
&  \del{p_g}{t} +
u \del{p_g}{x} + \gamma_{g} p_g \del{u}{x} 
=  (\gamma_{g} - 1) \alpha \del{u}{x}, 
\label{eq:hd_gerg}\\
& \del{p_c}{t} +
u \del{p_c}{x} + \gamma_c p_c \del{u}{x} -
\del{}{x} \left( \kappa \del{p_c}{x} \right)
= - (\gamma_c - 1) \alpha \del{u}{x}, \label{eq:hd_pcerg}
\end{align}
where $\rho, u, p_g$ denote the density, flow
velocity, and pressure of the thermal component. The CR pressure $p_c$ defined
by the moment of the (isotropic part of) CR distribution function $f(p)$, 
\begin{align}
 p_{c}  = \frac{4 \pi}{3}\int v p^{3} f(p) dp .
\end{align}
evolves according to the equation \eqref{eq:hd_pcerg} that includes the
convection, adiabatic heating, and spatial diffusion with its coefficient
denoted by $\kappa$. Throughout this study, $\gamma_{g} = 5/3$ and $\gamma_{c} =
4/3$ are assumed for the specific heat ratios of the background plasma and
CRs, respectively.

The equation \eqref{eq:hd_pcerg} may be derived by taking the appropriate
moment of the diffusion-convection equation for CRs
\citep[e.g.,][]{1975MNRAS.172..557S}
\begin{align}
\del{f}{t} + u \del{f}{x} - \del{}{x} \left(\kappa'(p) \del{f}{x}\right)
= \frac{1}{3} \del{u}{x} p \del{f}{p}. \label{eq:dif-conv}
\end{align}
An arbitrary energy dependence of the diffusion coefficient $\kappa' (p)$ in
the original equation is eliminated and $\kappa$ roughly corresponds to that
of particles in the energy range containing most of the CR pressure. In the
present study, we also assume that $\kappa$ is constant both in space and time
for simplicity.

The advantage of the two-fluid model is its simplicity making it possible to
investigate the property of the system analytically. Extension to
magnetohydrodynamics (MHD) \citep{1986A&A...160..335W}, radiative shocks
\citep{2006A&A...452..763W}, and a model with the effect of an acoustic
instability \citep{2007A&A...463..195W,2009ApJ...690.1412W} were also
proposed. On the other hand, the most crucial difference of the two-fluid
model from 
kinetic models is probably the absence of the maximum CR energy,
which introduces differences in the steady-state solutions. However, we
believe that it will not affect the stability property of the system (see
discussion in Section \ref{sec:sum} for details).

\subsection{Injection Model}\label{sec:inje_model}

We adopt the injection model proposed by \cite{1993ApJ...406...67Z} based on
the idea of thermal leakage.  It defines the momentum boundary $p_{0}$ above
which particles are considered to be CRs and their transport obeys the
diffusion-convection equation \eqref{eq:dif-conv}. Namely, heating of the gas
component injects a fraction of thermal particles into CRs. Under the
assumption, we can obtain the 
equation \eqref{eq:hd_pcerg} by integrating the equation
\eqref{eq:dif-conv} above $p_{0}$ in momentum space. The particle injection
term appears because of this lower limit of integration. The injection
parameter $\alpha$ defined as
\begin{align}
\alpha = \frac{4\pi}{3} E(p_{0}) p_{0}^{3} f(p_{0}), \label{eq:alpha}
\end{align}
represents the energy density of the injected particle flux. 
Since the
particle injection term is written as a product of $\alpha$ and the spatial
gradient of the flow,
\begin{align}
S = \alpha \del{u}{x},
\end{align}
the injection at the subshock is dominant over the precursor. Notice that the
parameter $\alpha$ must be a function of both space and time because it is a
quantity determined by local density and temperature of the thermal
plasma. \cite{1993ApJ...406...67Z} and \cite{1994ApJ...424..263D}, however,
assumed that it is constant to make the problem analytically tractable.

In numerical simulations, we can easily calculate $\alpha$ more rigorously for
a given momentum boundary $p_0$ by assuming a distribution function of the
background plasma $f_{th}(p)$. For this purpose, we adopt the
(non-relativistic) Maxwellian distribution
\begin{align}
f_{th}(p) = n \left(\frac{1}{2\pi m k_{B} T }\right)^{3/2} \exp{\left[
-\frac{p^{2}}{2 m k_{B} T} \right]},
\end{align}
where $m, n, T$ are the proton mass, density and temperature of the background
plasma and $k_B$ denotes the Boltzmann constant, respectively. The parameter
$\alpha$ can then be written as follows,
\begin{align}
\alpha (p_0) = \frac{4 \pi}{3} E (p_0) p_0^3 f_{th}(p_0).
\label{eq:general_alpha}
\end{align}
The particle kinetic energy is given in the
relativistic form $E(p) = \sqrt{1 + (p/mc)^2} - 1$, where $c$ is the speed
of light. 
In the present study, the
injection model given by the equation \eqref{eq:general_alpha} is referred
to as self-consistent.

Note that the injection momentum $p_{0}$ is typically chosen to be a few times
the downstream thermal momentum $p_{th} = 2 \sqrt{m k_B T_{down}}$. This
choice is motivated by the fact that suprathermal particles in the downstream
region leaking out toward the upstream can be a seed population to the
acceleration process. The most important feature of the self-consistent
injection model is that the injection efficiency is regulated in response to
the downstream temperature changes due to the dynamical shock modification. One
can expect that the increase in CR pressure tends to reduce the subshock
strength and thus the injection efficiency and vice versa. Such a
self-consistent regulation of the injection, albeit simplified, takes into
account the feedback effect at least qualitatively.

In this study, we investigate both cases; the constant-$\alpha$ injection
(Section \ref{sec:inje_stab}) and the self-consistent injection (Section
\ref{sec:nonuni_inje}) to clarify the role of injection on the stability of
CRMSs.

\subsection{Analytical Solutions} \label{sec:anal_model}
Analytical steady-state CRMS solutions to the equations
\eqref{eq:hd_mass}-\eqref{eq:hd_pcerg} for a constant $\alpha$ were obtained
by \cite{1993ApJ...406...67Z}, which are the extension of the non-injection case
$\alpha=0$ originally given by \cite{1981ApJ...248..344D}. We here briefly
review the basic characteristics of these solutions.

\begin{figure}[tb]
 \includegraphics[width=\columnwidth]{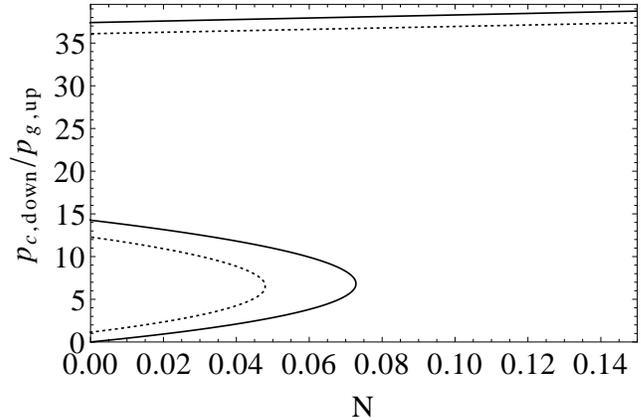}
 \caption{
Downstream CR pressure $p_{c}$ as a function of the upstream CR fraction $N$.
The solid and dashed lines show solutions for $M=6.5$, $\alpha=0$, and
 $M^{*}=6.5$, $\alpha/p_{g,up}=0.1$, respectively.
}
 \label{fig:RH}
\end{figure}

The solution in the non-injection case, in which acceleration of pre-existing
CRs is considered, is characterized by the Mach number $M$ and the fraction of
CRs $N$ in the far upstream
\begin{eqnarray}
 M=\frac{u}{C_{s}}, \\
 N=\frac{p_{c}}{p_{g}+p_{c}},
\end{eqnarray}
where $C_{s} = \sqrt{\gamma_g p_g / \rho}$ is the sound speed of the
background plasma. The solid line in \figref{RH} 
shows the relation between $N$ and the
downstream CR pressure $p_{c, down}$ for an upstream Mach number of 
$M = 6.5$. 
One immediately finds that, for $N \lesssim 0.07$, multiple solutions
exist for a given upstream state. This is a distinct feature for the system
absent in the hydrodynamic shock. For convenience, we shall call these
solutions, ``efficient'', ``intermediate'', and ``inefficient'' from the top
to bottom as shown in \figref{RH}, as they are characterized by CR
production efficiencies. The inefficient branch essentially corresponds to the
test particle limit and the modification is of only minor importance. On the
other hand, CRs absorb most of the kinetic energy in the efficient branch. The
substantial difference in the CR production efficiency, more than one order of
magnitude between the two in this particular case, motivates us to investigate
the stability of the multiple solutions.

Note that the subshock appears only in a relatively low CR fraction $N$ and
Mach number $M$ in the two-fluid model. 
For sufficiently large values of $N$ and/or $M$, the subshock 
eventually disappears and the smooth
transition connects quantities between the upstream and downstream. 
The absence of the
subshock may be, however, an artifact of the two-fluid model. It has been shown
that the subshock always exists in a fully kinetic treatment
\citep{2001RPPh...64..429M}. We thus concentrate our discussion on the
solution involving the subshock.

The basic feature does not change even when the injection is taken into
account. The dashed line of \figref{RH}
 shows the same diagram for the injection case
with a constant $\alpha / p_{g,up} = 0.1$ for a
modified upstream Mach number $M^{*} = 6.5$. The modified Mach number $M^{*} =
u/C_{s}^{*}$ is defined in terms of the sound speed
\begin{align}
C_{s}^{*} =
\sqrt{
\frac{\gamma_{g} p_{g}}{\rho}
\left(
1 - \frac{\gamma_{g}-1}{\gamma_{g}} \frac{\alpha}{p_{g}}
\right)}, \label{eq:modi_cs}
\end{align}
modified by the effect of injection. The structure of the solution is
essentially the same as the non-injection case. It may be seen that the range
of parameter $N$ where multiple solutions exist is somewhat narrower in the
injection case, which reflects the role of injection; i.e., it effectively
increases the CR pressure. According to \cite{1993ApJ...406...67Z}, there are
solutions involving not only a precursor, but also a {\it postcursor} behind
the subshock which is not seen in the non-injection case. However, we do not
consider such solutions in the present paper for simplicity and focus on the
stability of multiple solutions.

%%% Local Variables:
%%% mode: yatex
%%% TeX-master: "main.tex"
%%% End:
%#!platex main.tex

\section{STABILITY OF GLOBAL SHOCK STRUCTURE} \label{sec:1d_hd_stability}

The stability of the global structure of  the steady-state solutions of CRMS 
is investigated by direct
time-dependent numerical simulations of the 1D two-fluid equations
\eqref{eq:hd_mass}-\eqref{eq:hd_pcerg}. 
As for the numerical method, we adopt a
splitting method \citep{1987ApJ...315..385D} in solving the equations
\eqref{eq:hd_mass}-\eqref{eq:hd_pcerg}. Namely, we split the time step into a
diffusion phase and a non-diffusion phase. In the diffusion phase, the
following equation is solved (here a constant diffusion coefficient $\kappa$ is
assumed), 
\begin{align}
 \del{p_c}{t} = \kappa \deld{p_c}{x},
\end{align}
in an implicit manner using the Bi-CGSTAB method \citep{ISI:A1992HE40000012}
to update the CR pressure to $p_{c}^{*}$. In the non-diffusion phase, we solve
the equations 
\eqref{eq:hd_mass}-\eqref{eq:hd_pcerg} without the diffusion term by the
modified Lax-Wendroff method \citep{1967JCoPh...2..178R}, 
which has the second-order accuracy both in time and space, 
using $p_{c}^{*}$ updated in the diffusion phase.
For the CFL condition, we adopt 
a variable time step such that 
$\Delta t = 0.1  \Delta x / 
\mathrm{max}(u + (\gamma_{g}p_{g}+\gamma_{c}p_{c} /\rho)^{1/2}) $, 
where $\Delta x$ is the grid spacing, and $\mathrm{max}()$ indicates the
maximum value in the simulation box.  

The number of grids is set to be $N_{x}=5000$,
which we believe is sufficient for the following reasons.
\cite{1994ApJS...90..975F,1995ApJ...441..629F} concluded that
their numerical solutions of MHD-CRMSs well converge to analytical ones when
sufficiently high resolution is used $n_r \gtrsim 10-20$, where $n_{r}$ is 
defined as $n_r = \kappa / (u_s \Delta x)$ (where $u_s$ is the shock
speed). 
In the present
paper, the parameter is always chosen to be $n_r > 100$, sufficient to give
numerical solutions with reasonable accuracy and discuss the stability of the
analytical solutions.
We employ the fixed boundary
at the left-hand (upstream) side and the free boundary ($\partial/\partial x =
0$) at the right-hand (downstream) side of the box.
We have checked that the boundary conditions do not influence our numerical
results by enlarging the simulation domain by five times. 
Space and time are respectively normalized to the diffusion length 
$\kappa/u_{up}$ and the diffusion time
$\kappa/u_{up}^{2}$. 
Note that our simulations
are conducted in the shock-frame, so $u_{up} \sim u_{s}$.

\subsection{Non-Injection Case ($\alpha=0$)} \label{sec:noninje_stab}

We choose an analytical steady-state solution as an initial condition for the
time-dependent simulation to investigate the stability. While we do not put
any perturbations into the simulation, it evolves from those caused by
numerical errors mainly at the subshock inherent in any finite difference
schemes.

\begin{figure}[tb]
\centering
 \includegraphics[width=0.9\columnwidth]{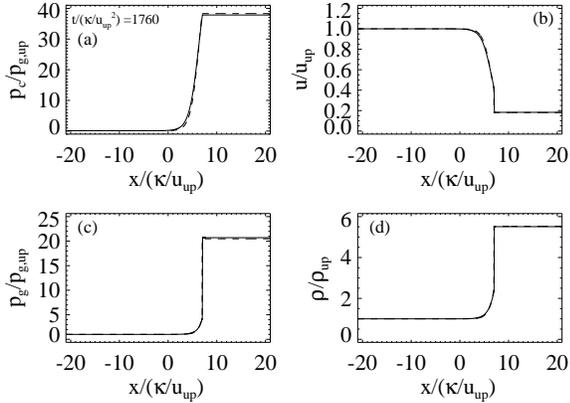}
 \caption{
Numerical solution for a CRMS with $M=6.5$ and $N=0.1$ where only one
 solution exists.
Normalized profiles of (a) the CR pressure, 
(b) background plasma flow velocity, (c)
 background plasma pressure, and (d) background plasma density are shown.
$p_{g,up}$, $u_{up}$, and $\rho_{up}$ are the upstream background plasma
 pressure, flow velocity, and density, respectively.
The initial and final states ($t/(\kappa/u_{up}^{2})=1760$) are shown in
 dashed and solid lines respectively.
}
 \label{fig:m6.5_n0.1_a0_para_t0-5}
\end{figure}

In this section, we study the non-injection case $\alpha = 0$ corresponding to
\cite{1981ApJ...248..344D}. 
Figures \ref{fig:m6.5_n0.1_a0_para_t0-5}(a)-(d)
show the results 
for $N=0.1$ and $M = 6.5$ in which only one solution involving a subshock
exists. In these kind of simulations, we have found that a numerical solution
always evolves into a steady state from which no appreciable changes are
observed, which is then regarded as the final state. One finds that the final
state (solid line) is almost unchanged from the initial condition (dashed
line), suggesting that the solution is stable.

\begin{figure}[tb]
 \centering
 \includegraphics[width=0.9\columnwidth]{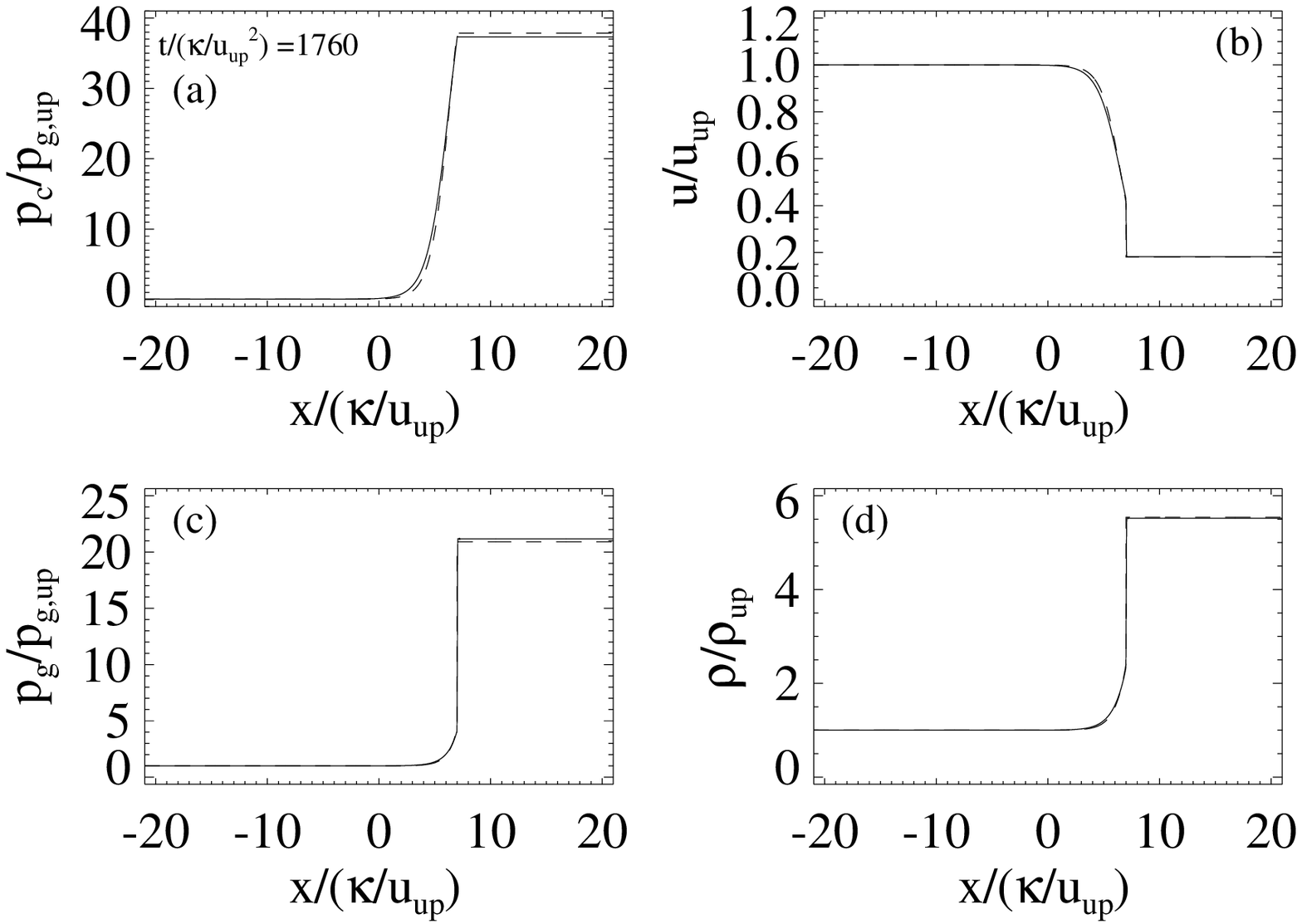}
 \caption{
Numerical solution for the efficient branch at
 $t/(\kappa/u_{up}^{2})=1760$ ($M=6.5$ and $N=0.05$).
The format is the same as Figure \ref{fig:m6.5_n0.1_a0_para_t0-5}.
}
  \label{fig:m6.5_n0.05h_a0_para}
\end{figure}

\begin{figure}[tb]
 \centering
 \includegraphics[width=0.9\columnwidth]{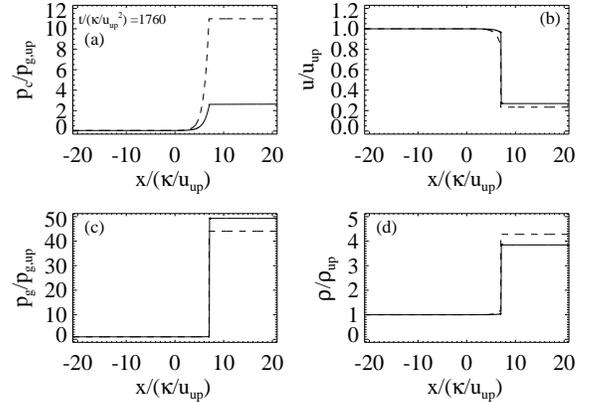}
 \caption{
Numerical solution for the intermediate branch at
 $t/(\kappa/u_{up}^{2})=1760$ ($M=6.5$ and $N=0.05$).
The format is the same as Figure \ref{fig:m6.5_n0.1_a0_para_t0-5}.
}
 \label{fig:m6.5_n0.05m_a0_para}
\end{figure}

\begin{figure}[tb]
 \centering
 \includegraphics[width=0.9\columnwidth]{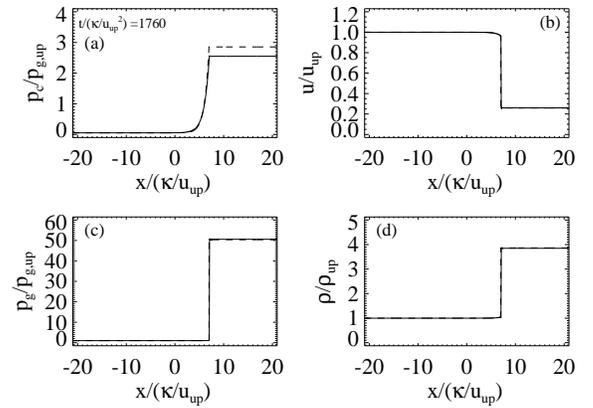}
 \caption{
Numerical solution for the inefficient branch at
 $t/(\kappa/u_{up}^{2})=1760$ ($M=6.5$ and $N=0.05$).
The format is the same as Figure \ref{fig:m6.5_n0.1_a0_para_t0-5}.
}
  \label{fig:m6.5_n0.05l_a0_para}
\end{figure}

Figures \ref{fig:m6.5_n0.05h_a0_para} - \ref{fig:m6.5_n0.05l_a0_para}
compare the results of three analytical
solutions corresponding to the efficient, intermediate, inefficient branches
found for $N=0.05$, respectively. 
We see that the downstream CR pressures of the efficient
and inefficient branches appear to be almost unchanged (Figures 
\ref{fig:m6.5_n0.05h_a0_para}(a)-(d) and \ref{fig:m6.5_n0.05l_a0_para}(a)-(d)), 
while that of the
intermediate branch decreases significantly
(Figures \ref{fig:m6.5_n0.05m_a0_para}(a)-(d)). 
This result indicates that the
intermediate branch is unstable while the others are stable. Note that the
difference between the background plasma parameters of the initial and final
states is relatively minor compared to the CR pressure for the simulation
started from the intermediate branch. We find that the final state in this
case corresponds to the inefficient solution. The reason for the minor
difference in the background plasma parameters is that the shocks of both the
initial and final states are intrinsically weakly modified ones.
Strictly speaking, the downstream CR pressure of the inefficient branch shows
a slight decrease, which we think is numerical. 
As we mentioned earlier,
we have checked the
convergence of numerical solutions to the analytical ones by increasing the
resolution.

\begin{figure*}[htb]
 \centering
 \subfigure{
 \includegraphics[width=0.9\columnwidth]{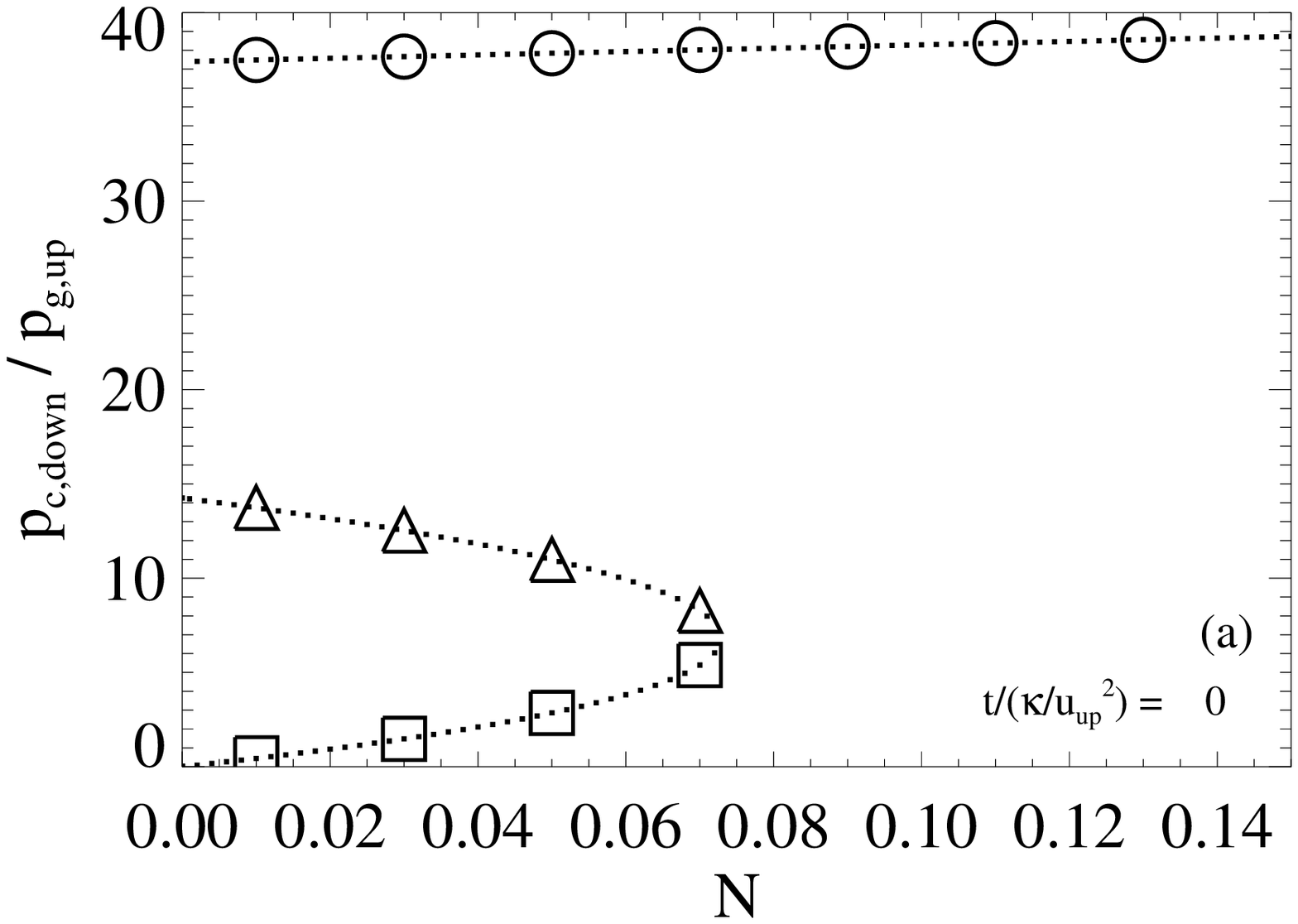}
 \label{fig:m6.5_a0_n-pc_t0}
 }
 \subfigure{
 \includegraphics[width=0.9\columnwidth]{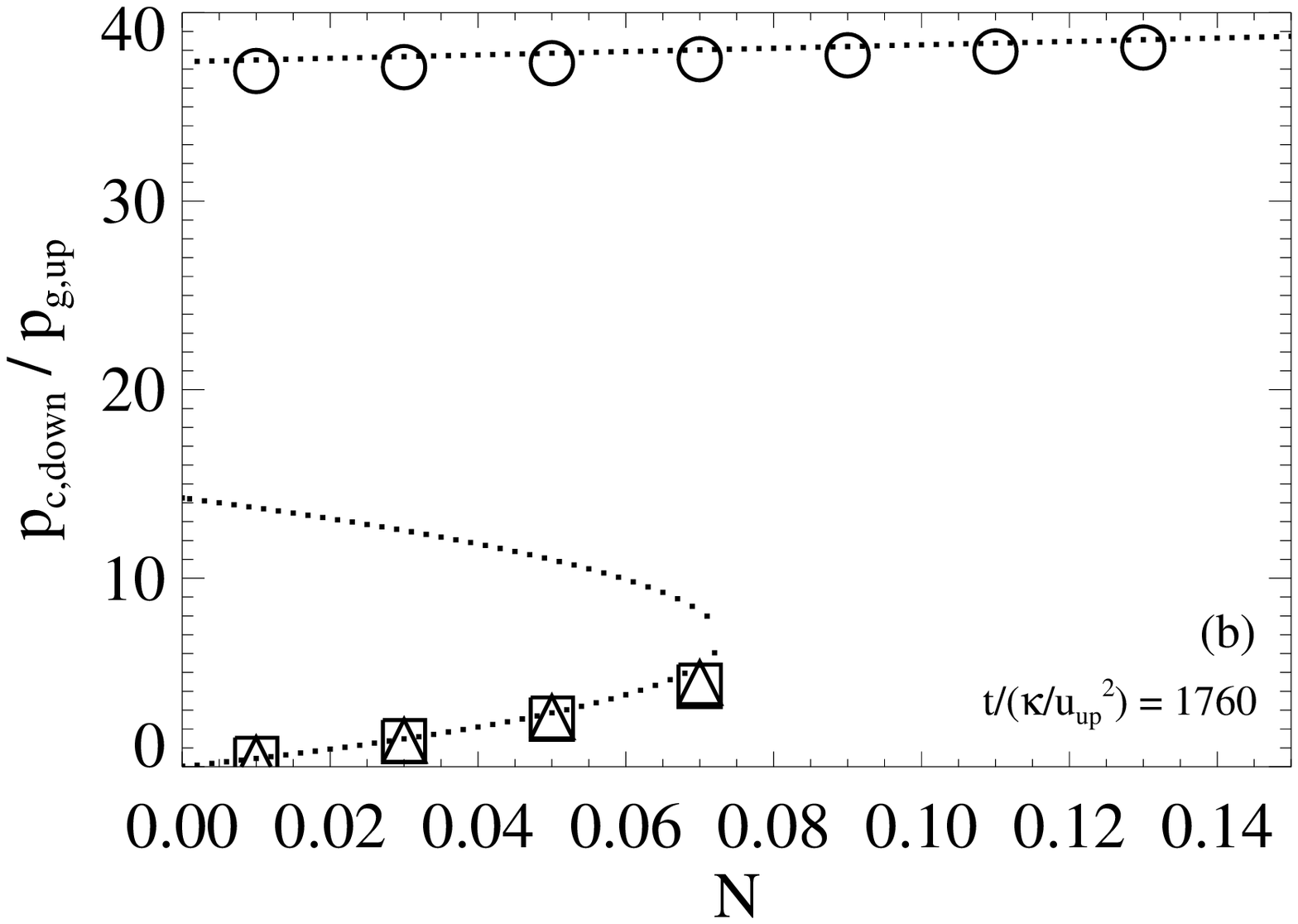}
 \label{fig:m6.5_a0_n-pc_t5.0}
 }

 \caption{
Summary of 
simulation results, for the non-injection case $M=6.5$ and 
$\alpha/p_{g,up}=0$ (constant) at (a) the initial state and (b) the final
 state.
The dotted line indicates the analytical steady-state solution.
Open circle, open triangle and open square show the results for the
 efficient, intermediate and inefficient branches respectively.
Simulations were conducted at 
$N=0.01$, $0.03$, $0.05$, $0.07$, $0.09$ $0.11$, and $0.13$.
}
 \label{fig:m6.5_a0_n-pc}
\end{figure*}

Figure \ref{fig:m6.5_a0_n-pc} summarizes 
the results for various initial conditions.
Each symbol represents a simulation run for a given upstream CR fraction $N$.
The downstream CR pressure, averaged over 250 grid points near the right-hand
side boundary, is shown in the vertical axis. In cases where there exists
multiple solutions for a given $N$, we investigate all the possibilities. The
initial conditions are indicated in (a), while the final states 
$t/(\kappa / u_{up}^{2}) = 1760$ are shown in (b). 
As was found in the case of $N=0.05$, the
efficient and inefficient branches exhibit only slight changes from the
initial conditions due to numerical errors as mentioned above. On the other
hand, the intermediate branch always shows the transition to the inefficient
branch.
This has been confirmed in the range $5 \le M \le 15$,
$0.01 \le N \le 0.13$, whenever multiple solutions exist.
The sampling intervals for $M$ and $N$ are $0.5$ and $0.02$, respectively.
Note that, for higher Mach numbers, there exists only one solution
(corresponding to the efficient state) in the range $N \ge 0.01$.

%% subfigure 
%%%%%%%%%%%%%%%%%%%%%%%%%%%%%%%%%%%%%%%%%%%%%%%%%%%%%%%%%%%%
\renewcommand*{\thesubfigure}{(\alph{subfigure})}
\renewcommand{\subfigcapskip}{-8pt}
%%%%%%%%%%%%%%%%%%%%%%%%%%%%%%%%%%%%%%%%%%%%%%%%%%%%%%%%%%%%

\subsection{Injection Case ($\alpha \neq 0$)}\label{sec:inje_stab}

We now study the effect of injection with a constant injection parameter
$\alpha$. As in the non-injection case, we can use the analytical solutions of
\cite{1993ApJ...406...67Z} presented in Section \ref{sec:anal_model} as the
initial conditions.

\begin{figure}[htb]
   \centering
 \includegraphics[width=0.9\columnwidth]{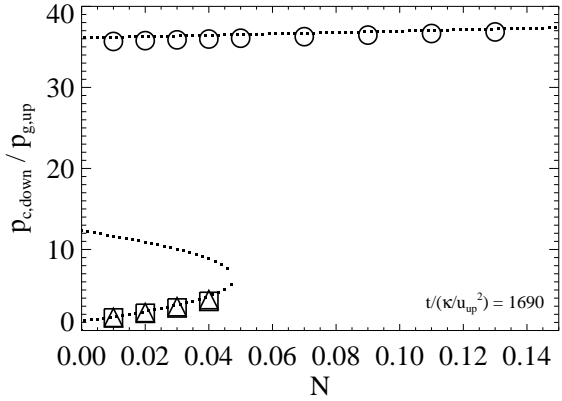}
 \caption{
Summary of 
simulation results for the injection case $M^{*}=6.5$ and 
 $\alpha/p_{g,up}=0.1$ (constant) at the final state.
The format is the same as \figref{m6.5_a0_n-pc}.
}
 \label{fig:m6.5_a0.1_n-pc}
\end{figure}

\figref{m6.5_a0.1_n-pc} shows the results with the same format as
\figref{m6.5_a0_n-pc} for $M^{*}=6.5$ and $\alpha/p_{g,up} = 0.1$. 
Note that the reason why the efficient branch in the injection case is
less efficient than that in the non-injection case (which may easily be
seen in Figure \ref{fig:RH}) is due to the definition of $M^{*}$ which is
a function of parameter $\alpha$.
One
immediately sees that the basic stability property is essentially unchanged,
i.e., the efficient and inefficient branches are stable while the intermediate
branch is always unstable and evolves into the inefficient one. Extensive
parameter survey in the
range $0.001 \le \alpha \le 1$, $5 \le M^{*} \le 15$ and $0.01 \le N \le
0.13$ again confirms that
the property does not change, although
the use
of different parameters modifies the structure of analytical solutions
itself 
(the sampling intervals are the same as previous ones for $M^{*}$ and $N$, 
and $0.001$, $0.01$, $0.1$, $1$ for $\alpha$). 
One might naively expect that the introduction of injection tends to
make the acceleration more efficient, but this is not the
case.

\subsection{Self-consistent Injection Case}\label{sec:nonuni_inje}

Unlike the case with a constant $\alpha$ (including non-injection case), no
analytical solution is known for the self-consistent injection case. However,
since the injection is the strongest at the subshock where the downstream
$\alpha$ plays an essential role, we initialize the simulation in the
following way. First, we set up an initial condition for the background plasma
parameters and CR pressure using an analytical solution for a constant
$\alpha$. We then calculate $p_0$ using the 
equation \eqref{eq:general_alpha} and
the downstream density and temperature $T$. This $p_0$ is kept constant during
the entire simulation. The parameter $\alpha$ can now be calculated by using
the local density and temperature, and thus becomes a function of both space
and time. Notice that the parameter $\alpha$ so calculated in the precursor
and upstream differs from the original value even at the initial
condition. The inconsistency due to this is, however, relatively minor as the
injection primarily occurs at the subshock, which is indeed confirmed by
simulation results discussed below.

\begin{figure}[htb]
 \centering
 \includegraphics[width=0.9\columnwidth]{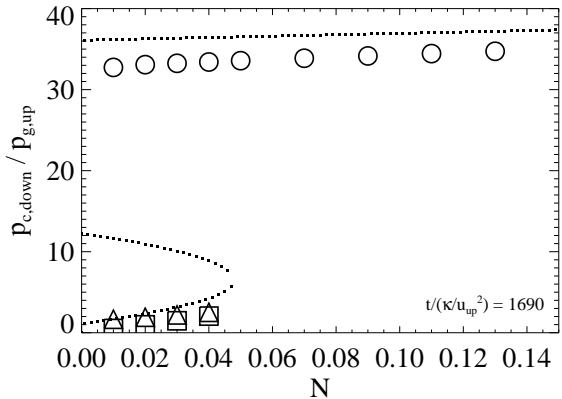}
 \caption{
Summary of simulation results for the self-consistent injection case 
$M^{*}=6.5$ at the final state.
The dotted line indicates the analytical steady-state solution for 
$\alpha/p_{g,up}=0.1$ (constant) for reference. 
The format is the same as \figref{m6.5_a0_n-pc}.
}
 \label{fig:m6.5_self-a0.1_n-pc}
\end{figure}

\figref{m6.5_self-a0.1_n-pc} shows the results with the self-consistent
injection with the same format shown in \figref{m6.5_a0.1_n-pc}. The initial
condition is set up by an analytical solution for a constant 
$\alpha/p_{g,up} = 0.1$. In each calculation, 
the momentum boundary $p_{0}$ differs slightly because we
set initial downstream $\alpha$ in all calculations to satisfy 
$\alpha/p_{g,up} = 0.1$ considering each different downstream state 
(e.g., $p_0/p_{th} \simeq 2.65$ for $N=0.13$ and
$p_0/p_{th} \simeq 2.75$ for $N=0.01$ of the inefficient branch). 
We see that the stability property is essentially
not affected by the different injection model. The only difference we can
find from \figref{m6.5_self-a0.1_n-pc} is that the solutions as a whole
slightly shift to lower CR pressure states from the initial condition
constructed for a constant $\alpha$. This may be explained by considering a
finite subshock width. Namely, since the injection flux is expressed by a
product of $\alpha$ and the flow divergence, the strongest injection occurs at
the subshock which is resolved by a finite number of grid points. The $\alpha$
parameter calculated by density and temperature in the subshock structure thus
gives an intermediate value between the upstream and downstream at which
the flow divergence is largest.
This means
that an effective $\alpha$ is somewhat smaller than the downstream value. We
have confirmed that the numerical solutions agree very well with analytical
solutions calculated using the effective $\alpha$ parameters 
evaluated from simulation results (assumed to be
constant). Therefore, the differences between the initial and final states are
injection model dependent. Such an issue is obviously beyond the scope of the
present study, and it should not be taken too seriously. It is rather
important to emphasize that the self-consistent injection does not introduce
appreciable differences to the stability of the CRMS solutions.

\subsection{Large-amplitude Perturbations}\label{sec:gas-shock}

So far we have investigated the stability against relatively small
perturbations caused by numerical errors, where the 
intermediate branch  is  always unstable and evolves into a
less efficient state. For application to realistic astrophysical situations
where the shock parameters may change in time (e.g., slowing down of SNR
shocks, inhomogeneous upstream media), 
it may also be important to understand the stability
property against large-amplitude perturbations.

\begin{figure}[htb]
 \centering
 \includegraphics[width=0.9\columnwidth]{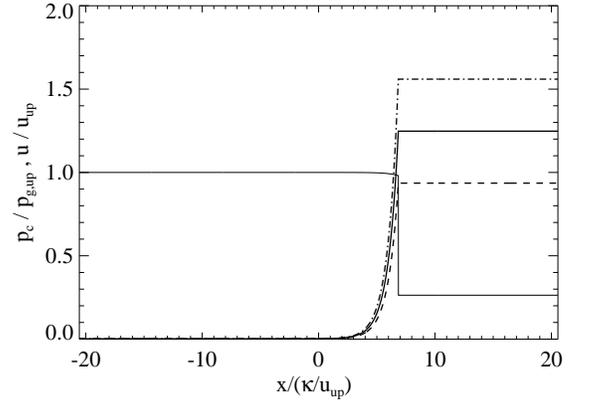}
 \caption{
Initial profiles of perturbed CR pressure $p_{c}$ for the inefficient
 solution with $M^{*}=6.5$, $N=0.002$, and $\alpha/p_{g,up}=0.1$.
The thin and thick solid lines show respectively the flow velocity and the
 CR pressure for the analytical solution. Perturbed CR pressure profiles
 by amount $\pm 25\%$ are shown in the dashed and dashed-dotted lines
 respectively. 
}
 \label{fig:m6.5_n0.002l_a0.1_pc_vx}
\end{figure}

\begin{figure*}[htb]
 \centering
 \subfigure{
 \includegraphics[width=0.9\columnwidth]{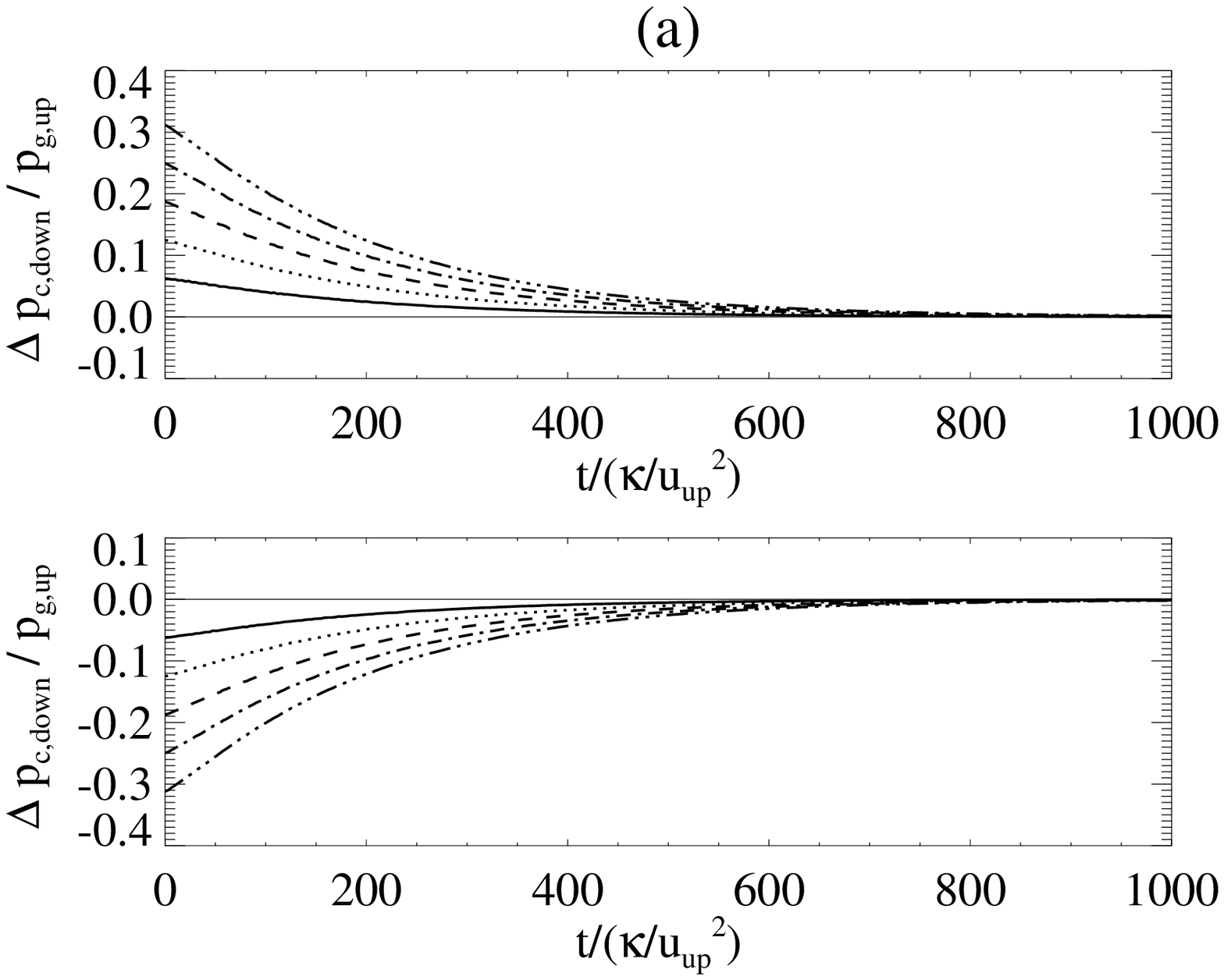}
 \label{fig:m6.5_n0.002l_a0.1_d-pc}
}
 \subfigure{
 \includegraphics[width=0.9\columnwidth]{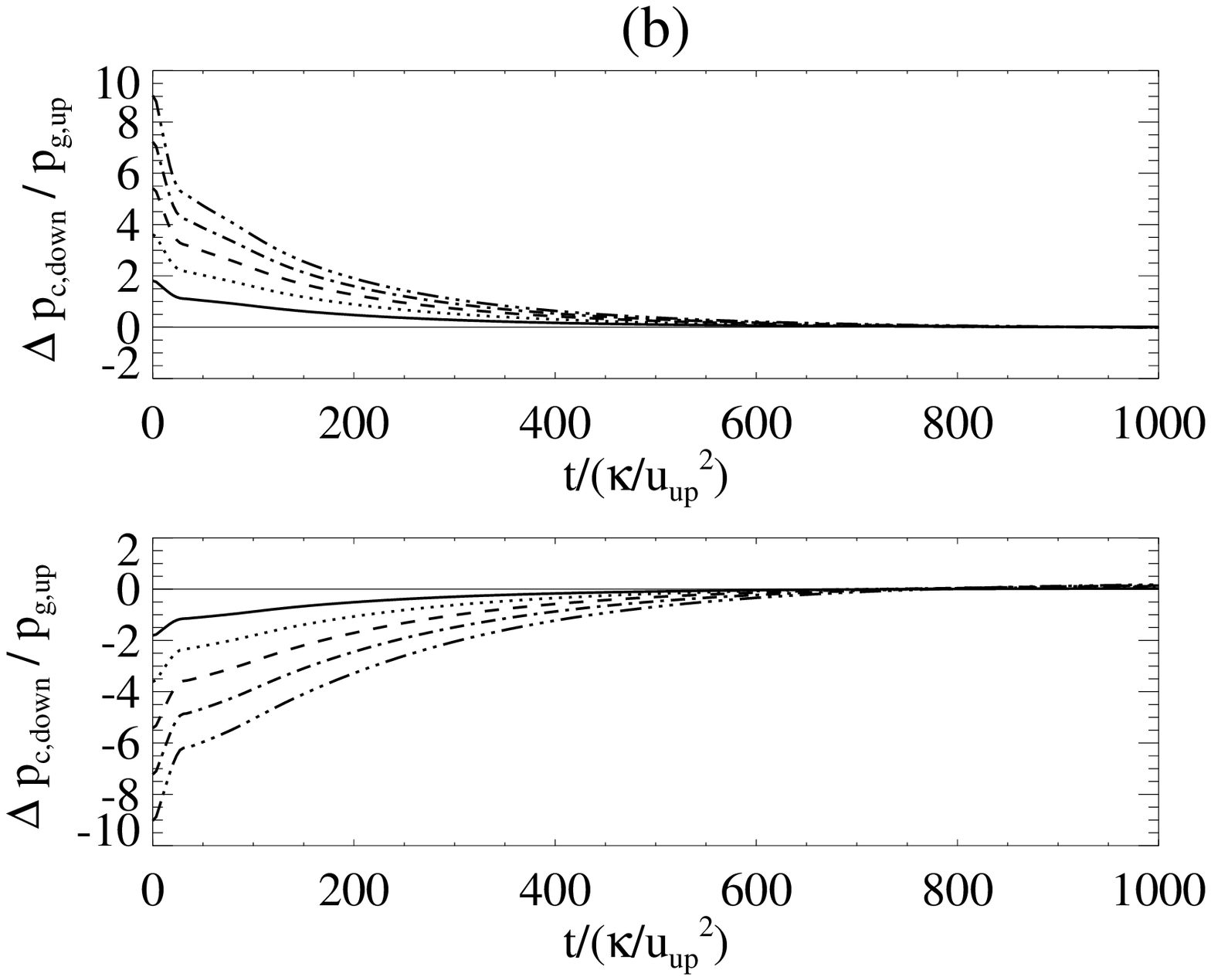}
 \label{fig:m6.5_n0.002h_a0.1_d-pc}
}

 \caption{
Time evolution of the deviation of downstream CR pressure from the
 unperturbed case for $M^{*}=6.5$, $N=0.002$ and $\alpha/p_{g,up}=0.1$.
The panels (a) and (b) show the inefficient and efficient branches
 respectively. 
In each case, positive and negative perturbation runs are shown in the top
 and bottom. The absolute amplitude of perturbation is shown with
 different line types (solid: $5 \%$, dotted: $10 \%$, dashed: $15 \%$, 
dashed-dotted: $20 \%$, dashed double-dotted: $25 \%$).
}
 \label{fig:m6.5_n0.002_a0.1_d-pc}
\end{figure*}

\begin{figure*}[htb]
 \centering
 \subfigure{
 \includegraphics[width=0.9\columnwidth]{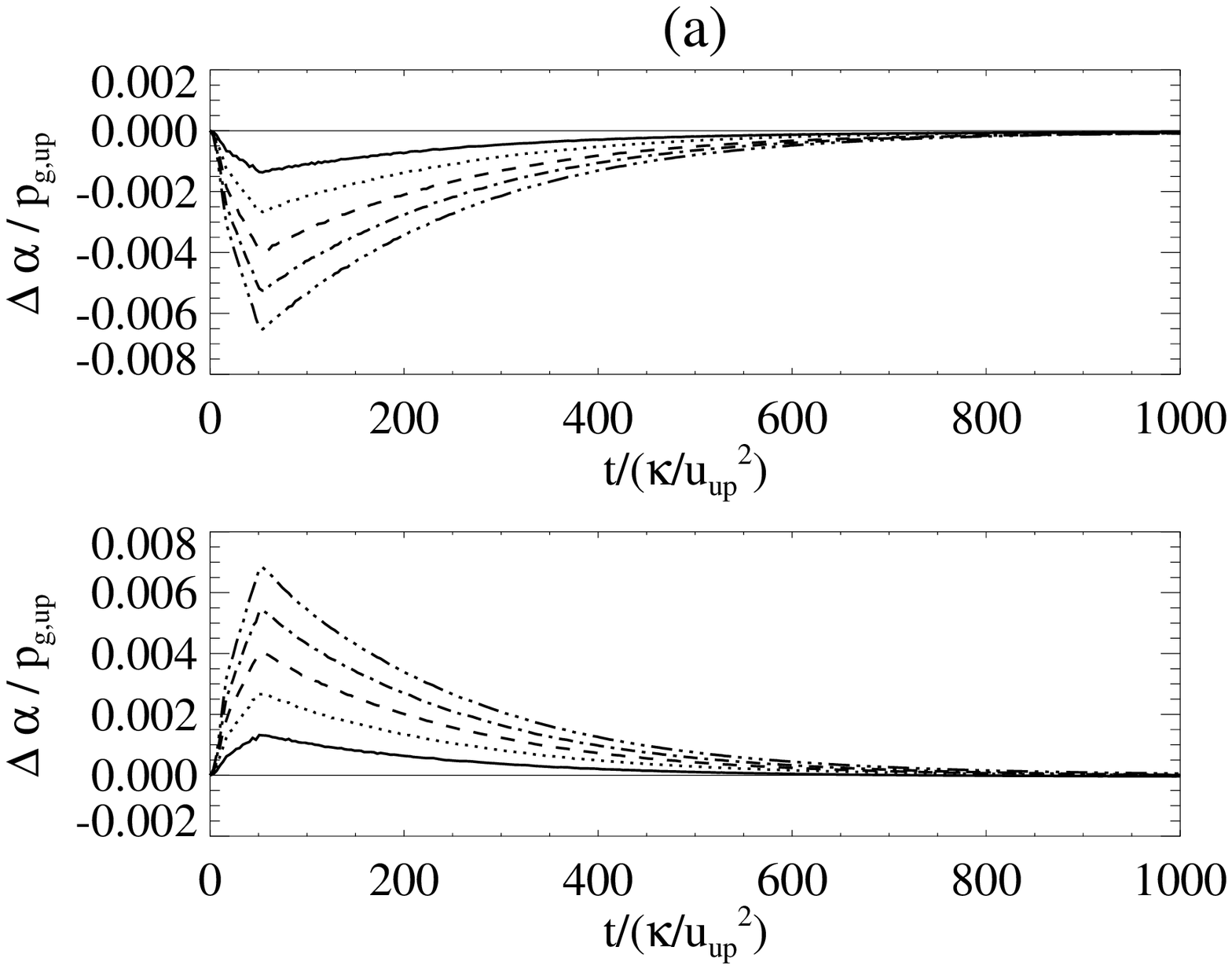}
 \label{fig:m6.5_n0.002l_a0.1_d-a}
}
 \subfigure{
 \includegraphics[width=0.9\columnwidth]{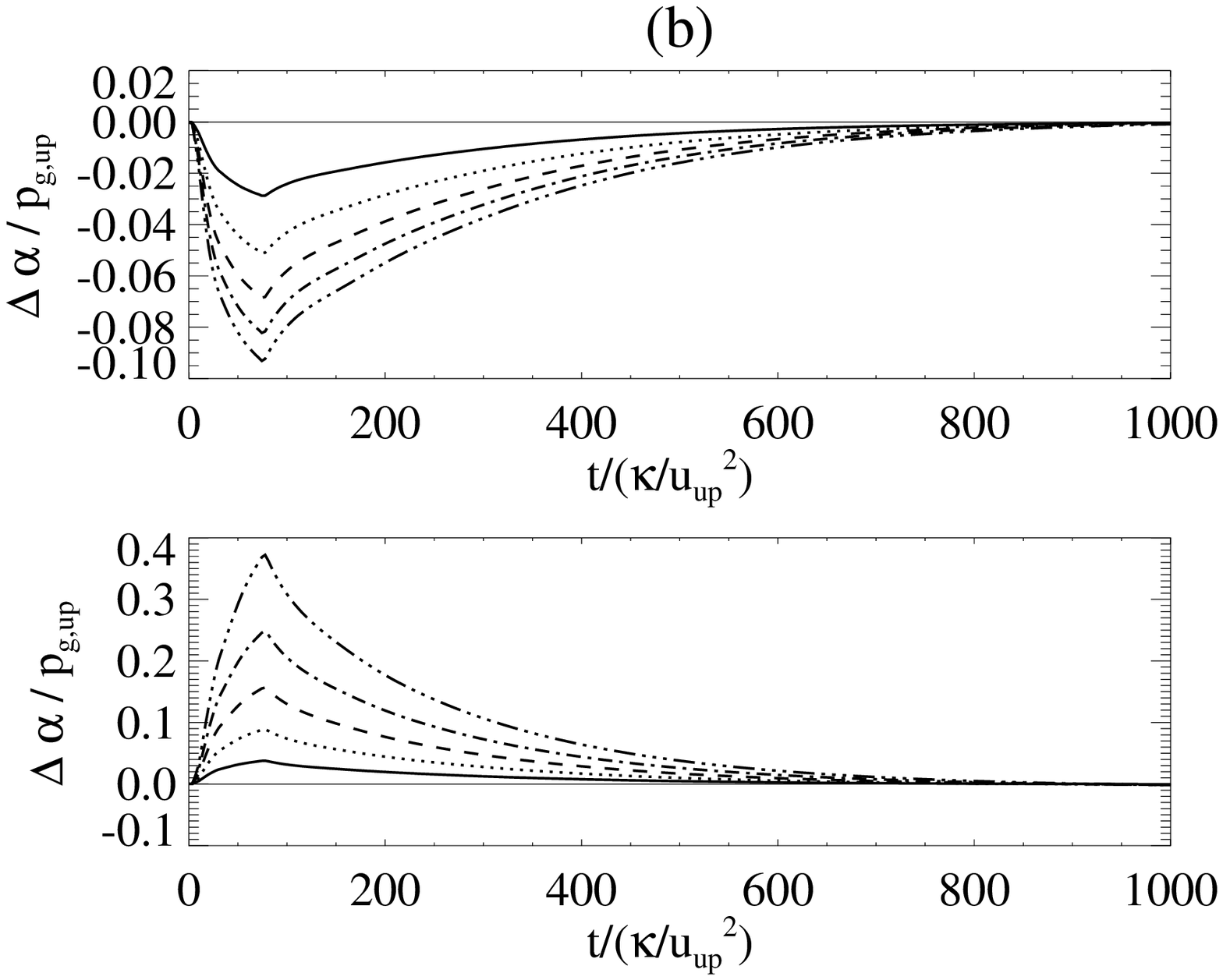}
 \label{fig:m6.5_n0.002h_a0.1_d-a}
}
 \caption{
The same as Figure \ref{fig:m6.5_n0.002_a0.1_d-pc} but for the downstream
 $\alpha$.
}
 \label{fig:m6.5_n0.002_a0.1_d-a}
\end{figure*}

We investigate the response of the system against large-amplitude
perturbations. Specifically, we change the downstream CR pressure 
$p_{c,down}$ at the
initial condition to investigate the behaviors in the $N-p_{c}$ diagram. 
Figure \ref{fig:m6.5_n0.002l_a0.1_pc_vx} shows an example of
perturbed and unperturbed profiles of the CR pressure. In order to obtain an 
initial perturbed profile $p_{c}^{'}(x)$, 
we multiply the analytical solution
$p_{c}(x)$ by a constant factor 
corresponding to the amplitude of perturbation.
On the other hand, hydrodynamic quantities $u$, $p_{g}$
and $\rho$ are remain  unchanged.
Figures \ref{fig:m6.5_n0.002_a0.1_d-pc} and
\ref{fig:m6.5_n0.002_a0.1_d-a} show the response of the system obtained by
numerical simulations for (a) the inefficient, 
and (b) the efficient branches, respectively.
We choose an analytical steady-state solution for
$M^{*}=6.5$, $N=0.002$, and a constant $\alpha/p_{g,up} = 0.1$ 
on which initial large-amplitude perturbations are
imposed. 
The results with perturbations up to $\pm 25 \%$ of the unperturbed
state are shown in these figures. The ratios of
the momentum boundary to the downstream thermal momentum for this case are
$p_0/p_{th} \simeq 2.35$ and $p_0/p_{th} \simeq 2.47$ for the inefficient and
the efficient branches respectively. The CR pressure in the downstream as well
as the $\alpha$ parameter shown in Figures \ref{fig:m6.5_n0.002_a0.1_d-pc} and
\ref{fig:m6.5_n0.002_a0.1_d-a} are calculated by taking the 
%average over the 
%downstream ($\simeq 90 \%$ of the region from the right-hand boundary). 
average over the values, respectively in all cells between the downstream
boundary and $\sim 10 \%$ inside the uniform region downstream.
We can
see that the injection parameter $\alpha$ immediately increases (decreases) in
response to the decrease (increase) in the CR pressure. This confirms the
feedback effect of injection due to dynamical modification of the
shock. Nevertheless, the simulation results show that the numerical solutions
quickly converge into the solution obtained without perturbations, suggesting
that these solutions are 
stable even against large-amplitude
 perturbations and the injection does not play a
role for modifying the stability.

%% subfigure
%%%%%%%%%%%%%%%%%%%%%%%%%%%%%%%%%%%%%%%%%%%%%%%%%%%%%%%%%%%%
\renewcommand*{\thesubfigure}{(\alph{subfigure})}
 \renewcommand{\subfigcapskip}{-8pt}  
%%%%%%%%%%%%%%%%%%%%%%%%%%%%%%%%%%%%%%%%%%%%%%%%%%%%%%%%%%%%

%% subfigure counter
%%%%%%%%%%%%%%%%%%%%%%%%%%%%%%%%%%%%%%%%%%%%%%%%%%%%%%%%%%%%
\renewcommand*{\thesubfigure}{(\alph{subfigure})}
\renewcommand{\subfigcapskip}{-8pt}  
%%%%%%%%%%%%%%%%%%%%%%%%%%%%%%%%%%%%%%%%%%%%%%%%%%%%%%%%%%%%

\begin{figure*}[htb]
 \centering
 \subfigure{
 \includegraphics[width=0.9\columnwidth]{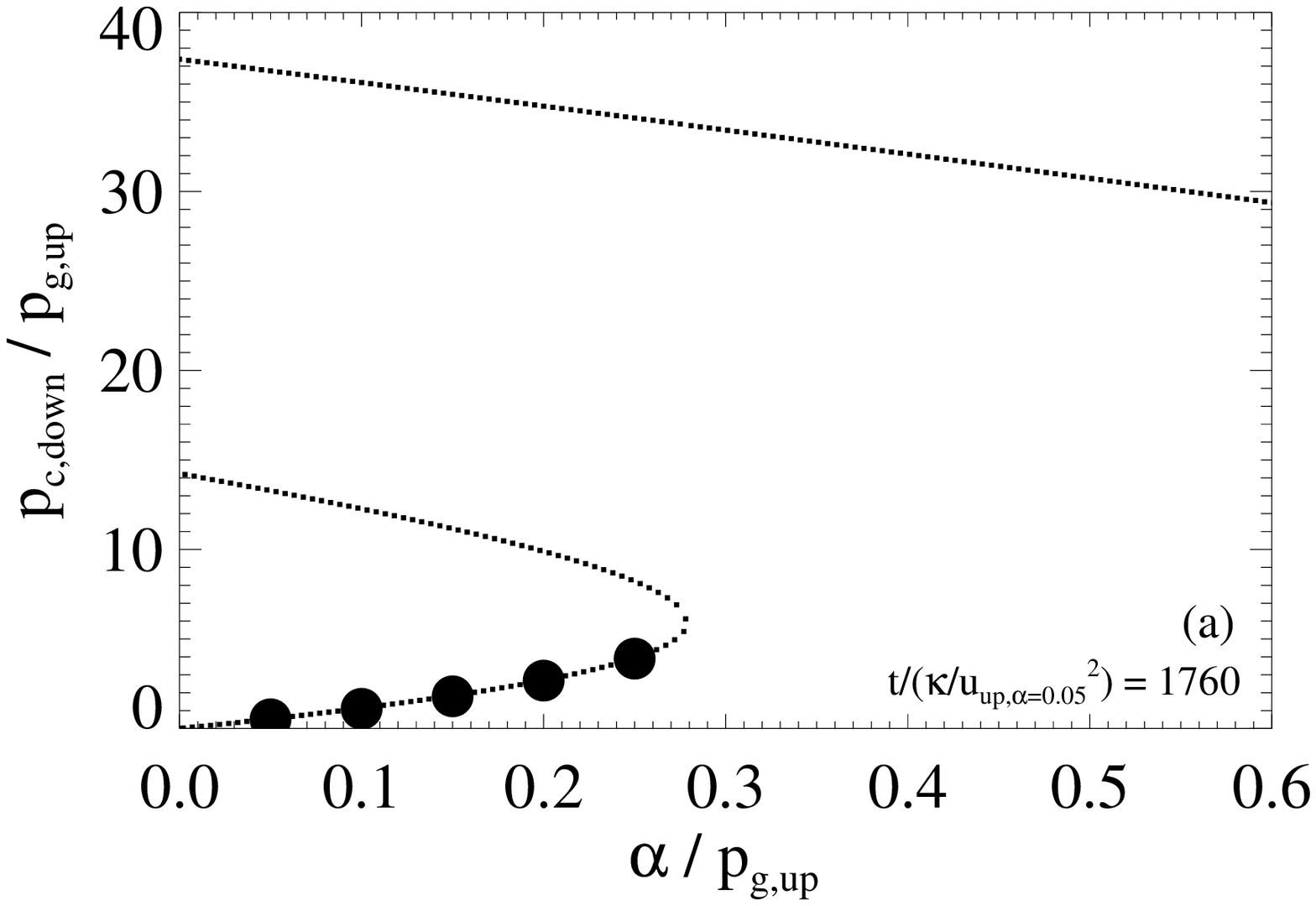}
  \label{fig:m6.5_n0_alpha-pc_t5}
}
 \subfigure{
  \includegraphics[width=0.9\columnwidth]{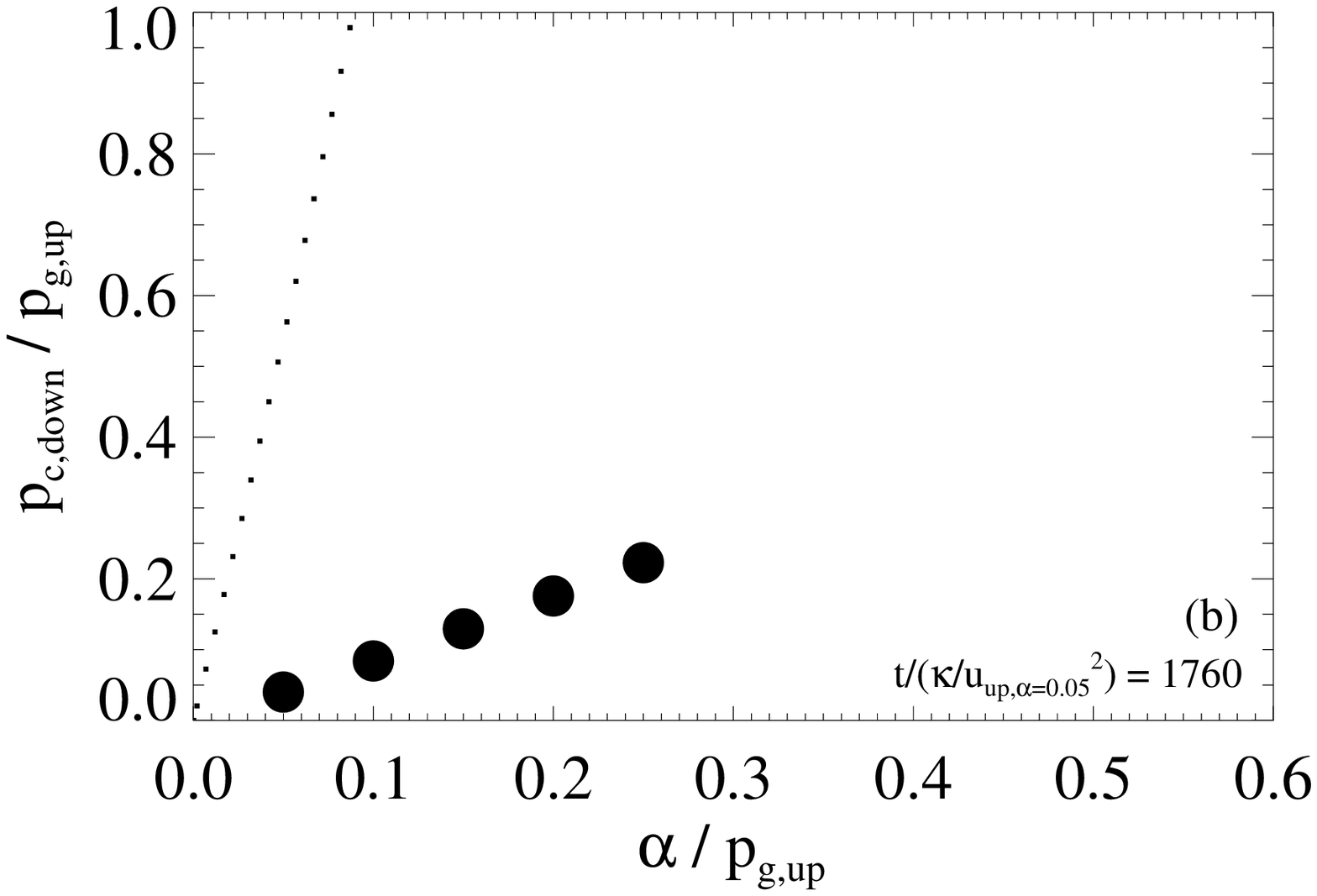}
  \label{fig:m6.5_n0_self-alpha-pc_t5}
}
 \caption{
Asymptotic states of simulations started from hydrodynamic shocks of
 $M=6.5$ with (a) constant $\alpha$ injection, (b) self-consistent
 $\alpha$ injection. 
The dotted lines indicate the analytical solution for $M^{*}=6.5$ with
 constant $\alpha$ for reference.
}
 \label{fig:m6.5_n0_alpha-pc}
\end{figure*}

We finally discuss numerical simulation results of time evolution from a 
hydrodynamic shock with injection.
This is particularly important in that it would be more or less
similar to the situation realized in a realistic astrophysical scenario.
\figref{m6.5_n0_alpha-pc} summarizes the 
results for shocks with a Mach
number of $M^{*}=6.5$ without pre-existing CRs ($N=0$), 
with (a) constant-$\alpha$ injection, and (b)
self-consistent injection.

For the constant $\alpha$ injection case shown in
\figref{m6.5_n0_alpha-pc}(a), we find that the final states of numerical
simulations with $0.05 \leq \alpha/p_{g,up} \leq 0.25$ 
settle into the inefficient branch of the analytical solutions shown by dotted
lines at $t/(\kappa/u_{up,\alpha=0.05}^{2})=1760$, where 
$u_{up,\alpha=0.05}$ is the upstream flow velocity of the background
plasma at  $\alpha/p_{g,up}=0.05$. As seen in the equation \eqref{eq:modi_cs}, 
the upstream sound velocity decreases with increase in  
the parameter $\alpha$ by 
a factor of $\sqrt{1-(2/5)(\alpha/p_{g,up})}$. 
Since the upstream flow velocity ($u_{up}$) depends on $\alpha$, 
we choose $u_{up,\alpha=0.05}$
as a representative value for the unit of time.
We have also checked the development
beyond this time but found no evidence for any further evolution, consistent
with the fact that the inefficient branch is stable against large-amplitude
perturbations. 
We also
conduct simulations with $\alpha/p_{g,up} = 0.3$ or even larger, where
only the efficient branch of solution exists. In these cases, the pressure
balance across the shock is broken because of strong modification of the
shock. As a result, the shock propagates toward upstream and the solution
settles into the efficient branch but with a different Mach number. Although
we are not able to plot the simulation result on \figref{m6.5_n0_alpha-pc}(a)
for this reason, it is certainly true that the time asymptotic state is on the
efficient branch. One might notice that the CR pressure of the analytical
solution on the efficient branch decreases as the injection
parameter $\alpha$ increases. 
This is because the plot is made for a fixed $M^{*}$ which
is a function of $\alpha$ and is therefore not surprising.

\figref{m6.5_n0_alpha-pc}(b) shows the final states of simulations with the
self-consistent injection (notice the different vertical scale). We choose
$p_{0}$ in such a way that an initial $\alpha$ determined by the downstream
background plasma density and pressure corresponds to 
$0.05 \leq \alpha/p_{g,up} \leq 0.25$. 
The analytical solution for a constant
$\alpha$ is also shown for reference. The simulation results always converge
to solutions below the reference solution. The reason for this is the same as
that given in Section \ref{sec:nonuni_inje} (i.e., due to a smaller effective
$\alpha$), and is not important. 

All these results indicate that the solutions on the efficient and inefficient
branches are stable even against large-amplitude 
perturbations, independent of the
assumption of the injection model. 
It is also worth mentioning the case with a finite upstream  CR fractions
($N>0$) at the initial state, which is more realistic for the 
astrophysical applications.
In such cases, we have confirmed that the time asymptotic states are 
also  on the inefficient branch if $N$ is relatively low so that  the 
inefficient branch exists.
Therefore, 
if one considers realistic time evolution of an astrophysical shock, 
the asymptotic state realized in nature will very 
likely to be the least efficient state in terms of particle acceleration for
given upstream parameters.

%%%%%%%%%%%%%%%%%%%%%%%%%%%%%%%%%%%%%%%%%%%%%%%%%%%%%%%%%%%%
\renewcommand*{\thesubfigure}{(\alph{subfigure})}
 \renewcommand{\subfigcapskip}{-8pt}  
%%%%%%%%%%%%%%%%%%%%%%%%%%%%%%%%%%%%%%%%%%%%%%%%%%%%%%%%%%%%

%%% Local Variables:
%%% mode: yatex
%%% TeX-master: "main.tex"
%%% End:
%#!platex main.tex

\section{SUMMARY AND DISCUSSION} \label{sec:sum}

In the present paper, we have investigated the stability of the global
structure of the CRMS by using
the two-fluid model with and without the effect of injection. The system is
known to have up to three distinct solutions in some regions in parameter
space, which are respectively referred to 
 as ``efficient'', ``intermediate'', and
``inefficient'' in terms of corresponding CR production
efficiencies. Understanding the stability of these solutions is crucial for
the application of nonlinear shock acceleration theory 
to astrophysical shocks. By
performing direct time-dependent numerical simulations, we have studied the
stability for the multiple solutions in a wide range of parameters space by
changing the Mach number $M (M^{*})$, the fraction of upstream pre-existing
CRs $N$, and the injection parameter $\alpha$. Our simulation results can be
summarized as follows.

Firstly, numerical simulations with three initial states given by the
analytical solutions of CRMSs demonstrate that the efficient and inefficient
branches are stable, while the intermediate solution always shifts toward the
inefficient branch. 
We have also confirmed this downward transition even if large-amplitude 
perturbations are imposed on the intermediate solution independent of 
the ``direction'' of perturbation. 
This result is consistent with the earlier conjecture of
the bistable feature suggested by \cite{1981ApJ...248..344D} even without
invoking the so-called corrugation mode known to be unstable in a 
multi-dimensional system \citep{1998A&A...332..385M}.

Secondly, the stability property does not depend on the injection model and
efficiency. We have investigated both the constant-$\alpha$ injection, as well
as the self-consistent injection in which $\alpha$ is determined by the
instantaneous density and temperature of the background plasma. In particular,
the self-consistent injection model implements the feedback effect due to
dynamical shock modification. Whereas the structure of steady-state solution
certainly depends on the injection, the stability is hardly affected even in
the case of self-consistent injection.

Thirdly, the efficient and inefficient branches are shown to be stable even
against large-amplitude 
perturbations, again regardless of the injection model. The
feedback effect of the self-consistent 
injection in response to large-amplitude perturbations to the
downstream state does not play a role in regulating the
stability. 
Consequently, a hydrodynamic shock with injection evolves into the  
inefficient branch 
whenever it exists as a result of self-consistent time development.
For the injection
parameter above a critical value in which only one solution corresponding to
the efficient branch exists, the shock structure drastically develops into the
strongly modified one. 
This suggests that the time asymptotic solution of the nonlinear shock is
likely to be the least efficient state for given parameters of the shock.

Our conclusions on the stability of the CRMS are based on the framework of the
two-fluid model. However, judging from the insusceptibility of the stability
property to otherwise important shock parameters ($M(M^{*})$, $N$,
$\alpha$ and the injection model), 
we believe that it will 
remain the same even in a fully kinetic treatment. The limit of the two-fluid
model has been discussed in the literature 
\citep{1990ApJ...353..149K,1991SSRv...58..259J,2001RPPh...64..429M}. 
It has
been suggested that the model gives essentially the correct description of the
CRMS provided that the adiabatic index of CRs $\gamma_c$ is adequately chosen
in the range $1 < \gamma_c < \gamma_g$. In the two-fluid model, increase of
$\gamma_c$ results in the shrinkage of the region of multiple solutions and
vice versa \citep[e.g.,][]{2001ApJ...546..429B}. 
The effective $\gamma_c$ in a kinetic model is determined by
solving self-consistently the modified shock structure.  The crucial
assumption in doing so is the maximum energy of CRs. Since the CRs absorb the
available kinetic energy through the positive feedback of shock modification,
the CR production rate tends to diverge and no steady-state solution would be
obtained unless one imposes a cut-off energy above which CRs escape from the
system. This makes the shock virtually {\it radiative} in the sense that the
effective $\gamma_c$ approaches  unity, which thus enlarges the region of
multiple solutions. 
%The variation of $\gamma_{c}$ influecnce the shape of energy spectrum
%because the total compression ratio of CRMSs depends on $\gamma_{c}$.
%When $\gamma_{c}$ approaches unity, the spectrum become harder.
%In addition to the simulations described in Section 3, 
We have also conducted 
simulations  with different $\gamma_{c}$, and confirmed that 
the bistable feature is insensitive to this parameter.
%These results imply that the introduction of
%self-consistently determined $\gamma_{c}$ in kinetic model do not less affect
%the bistable feature predicted from two-fluid model.  
In any case, solutions of the CRMS based on the kinetic
model have been obtained and confirmed  the existence of multiple
solutions \citep{1997ApJ...485..638M,1997ApJ...491..584M,2000ApJ...533L.171M,2004APh....21...45B,2005MNRAS.361..907B,2008MNRAS.385.1946A,2009ApJ...694..951R}. 
Rigorous proof of the stability in the kinetic regime is however left for
the future investigation.
Note that, in a kinetic model, multiple solutions seem to exist
for much higher Mach numbers, e.g., $M>100-1000$, 
which is not the case in the two-fluid model, 
probably due to the existence of the cut-off energy. The disappearance
of the subshock in the two-fluid model can also be explained similarly.

The fact that both the efficient and inefficient branches are stable even
against a large-amplitude 
perturbation makes it even more important to understand the
detailed structure of the CRMS solutions. More specifically, understanding the
critical parameters which distinguish the regions of single and multiple
solutions needs to be clarified for astrophysical applications. For instance,
considering realistic time evolution of a SNR shock propagating in the
interstellar medium, it may settle either on the inefficient or efficient
branches depending on the 
Mach number, upstream CR fraction, and injection rate. 
The physics of injection is still a
controversial issue and certainly beyond the scope of the present paper. The
injection is indeed determined as a result of thermalization involving
complicated physics of collisionless shocks. There exist plenty of theoretical
and numerical studies of injection processes, which indicates that the
injection processes and/or its efficiencies depend on the orientation of
magnetic fields, plasma $\beta$, and Mach numbers 
\citep[e.g.,][]{1995A&A...300..605M,2001JGR...10621657S,1990GeoRL..17.1821S,2010PhRvL.104r1102A}. 
The injection not only controls the number of particles accelerated by
the shock but also the total energy converted into CRs through nonlinear shock
modification, possibly leads to an abrupt ``phase transition''. This kind of
discontinuous transition may occur even for a fixed injection rate because of
intrinsic nonlinearity of the modified shock as suggested previously by
\cite{2001RPPh...64..429M}. 
Note that the CR production rate at SNR shocks is still uncertain 
\citep{2009Sci...325..719H,2000ApJ...543L..61H,2013arXiv1304.1261F}, 
but both the efficient and the inefficient solutions may in principle
applicable at present.
Although our limited knowledge of the physics of injection
and the maximum energy makes it difficult to state anything conclusive in
predicting observational consequences of astrophysical shocks,
our results suggest that 
an actual SNR shock may reside in the inefficient state, so that 
the CR production rate is lower than previously
discussed based on the strongly modified solutions.

Finally, we mention that the role of turbulent heating in the precursor may be
of great importance in regulating the efficient solution. Instabilities driven
by the CR gradient in the precursor may convert the CR energy into waves and
then lead to substantial nonadiabatic heating of the background plasma, which
would have non-negligible influence on the nonlinear acceleration process. It
will certainly weaken the modified shock and possibly even destroy the
efficient branch itself. Such physics beyond the framework of conventional
nonlinear shock acceleration theory must also be incorporated to elucidate
particle acceleration at astrophysical shocks.

%%% Local Variables:
%%% mode: yatex
%%% TeX-master: "main.tex"
%%% End:

\acknowledgments

We thank 
M. Fujimoto, T. Yokoyama, K. Shirakawa and K. Higashimori 
for the helpful suggestions and insightful discussions.
We are grateful to 
the proofreading/editing assistance from the GCOE program of the
University of Tokyo.

%%%%%%%
\bibliographystyle{apj}
\bibliography{myref}
%%%%%%%

\clearpage

\end{document}